\documentstyle[12pt,cmb]{article}

\def\pmb#1{\setbox0=\hbox{#1}%
  \kern0.00em\copy0\kern-\wd0
  \kern0.03em\copy0\kern-\wd0
  \kern0.00em\raise.02em\copy0\kern-\wd0
  \kern0.03em\raise.02em\copy0\kern-\wd0\box0 }

\input psfig.tex

\def\cobedmr{{\it COBE}-DMR$\;$}

\def\bC{{\bf C}}

\def\bG{{\pmb {$\Gamma$}}}

\def\bof{{\bf f}}

\def\bs{{\bf s}}
\def\bn{{\bf n}}
\def\bm{{\bf m}}
\def\bF{{\bf F}}

\def\bA{{\bf I}}
\def\bA{{\bf A}}

\def\bc{{\bf c}}

\def\bL{{\bf L}}

\def\bS{{\bf S}}
\def\bN{{\bf N}}
\def\bC{{\bf C}}

\begin{document}
\heading{COSMIC MICROWAVE BACKGROUND ANISOTROPY\\ 
         IN THE {\it COBE} DMR 4-YR SKY MAPS}

\author{Krzysztof M. G\'orski $^{1,}$$^{2,}$$^{3}$} 
       {$^{1}$ Hughes/STX, NASA/GSFC, Code 685, Greenbelt, MD 20771, USA.} 
       {$^{2}$ TAC, Juliane Maries Vej 30, 2100 Copenhagen \O,
       Denmark {\rm (present address)}.} 
       {$^{3}$ {\rm on leave from} Warsaw University Observatory, 
       Warsaw, Poland.}

\begin{abstract}{\baselineskip 0.4cm 
The {\it COBE} satellite has provided the only comprehensive
multi-frequency full-sky observations of the microwave sky
available today. Assessment of the observations requires a
detailed likelihood analysis to extract the maximum
amount of information present in the noisy data. I
present a specific method for estimating the CMB anisotropy
power spectrum independent of any assumptions about the
underlying cosmology, and then use standard image processing
techniques to generate the most revealing corresponding
maps of the signal. The consistency of the data at the three
available frequencies provides strong support to the
assertion that we are being provided with our first glimpse
of the last scattering surface.
}
\end{abstract}

\section{Introduction}

Operations of the Differential Microwave  Radiometer instruments
(DMR), the last active experiment 
on board NASA's COsmic Background Explorer ({\it COBE}) satellite, were terminated
in December 1993, concluding four years of measurements of anisotropy of the cosmic
microwave background (CMB) radiation. 
The final product of the DMR-team work --- full sky maps of the microwave sky
at 31.5, 53, and 90 GHz --- were released to the astronomical community in January 1996.
A generic description of the final DMR data set and brief summary of 
the results of
the DMR-team analysis  were given in \cite{b96}.
The issues of prime concern  in the DMR-team work included the following:

\noindent 1) modeling and removal of identified  
systematic artifacts from
the data (see \cite{ksys}) to produce  sky maps suitable for cosmological studies,

\noindent 2) studies of potential signal contamination by the Galaxy \cite{kgal}, 
and/or the extragalactic 
sources \cite{extrag},

\noindent 3) testing of the hypothesis of 
primordial origin of the measured CMB anisotropy \cite{rms}
by evaluation of the frequency dependence of the $(\delta T/T)_{rms}$ in the sky maps,

\noindent 4) testing of the hypothesis of gaussianity of statistics of the measured 
anisotropies \cite{kgauss},

\noindent 5) evaluation and analysis of the  auto- and cross-correlation functions 
of the sky maps \cite{hcorr},

\noindent 6) determination of the angular power spectrum of the CMB anisotropy
\cite{g96}, \cite{hsp}, \cite{w96}.  

\noindent Other analyses of the 4 year data presented so far include \cite{bw},
\cite{teg1}, and Bond and Jaffe in these Proceedings.

\begin{figure}[t]
\psfig{file=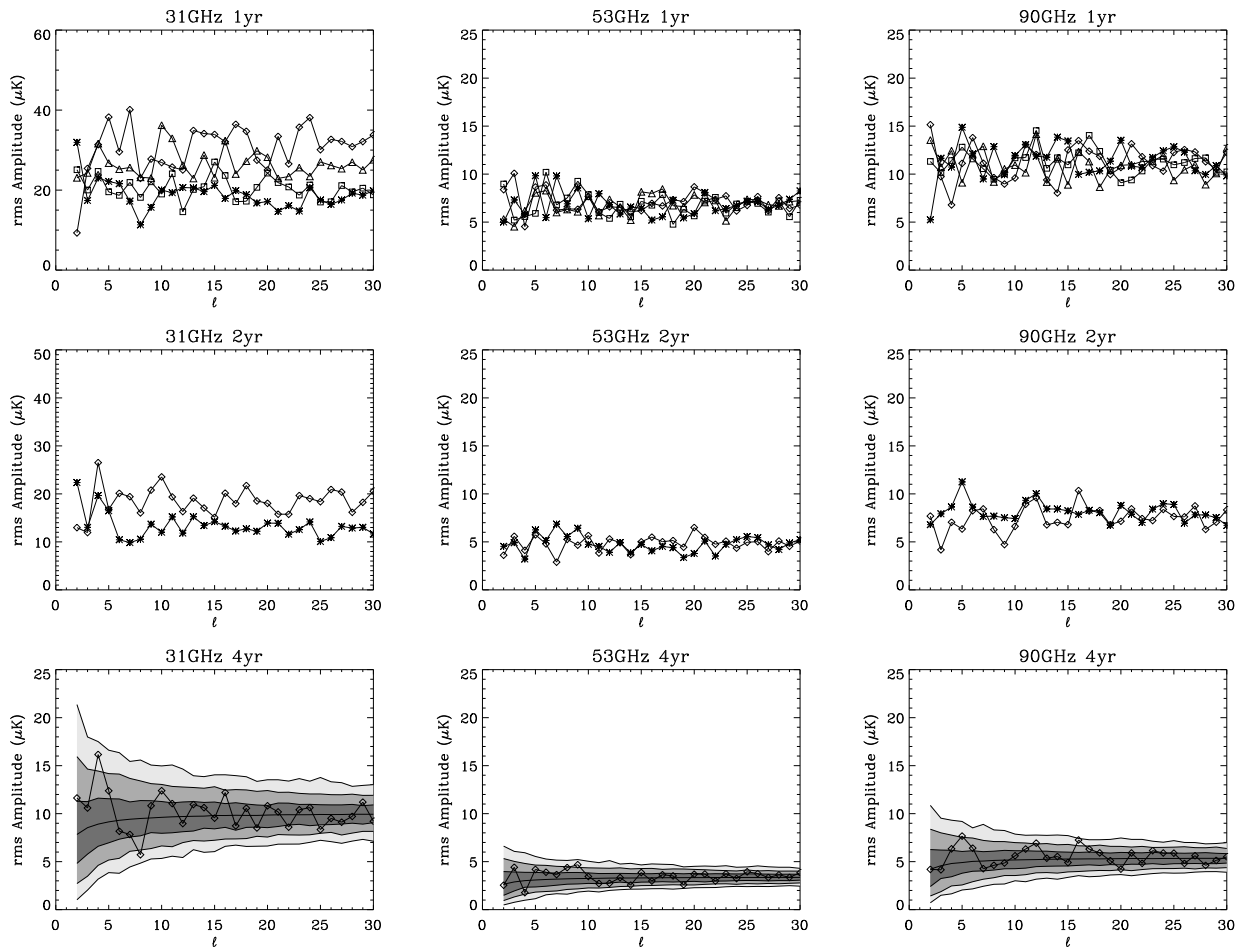,width=6.7in,angle=90}
\caption{Power spectra of the DMR (galactic frame, extended Galaxy cut,
3881 pixels) difference maps, (A-B)/2,
at each frequency of observations, 
for each yearly sky maps (top row, stars --- year one,
squares - year 2, triangles --- year 3, diamonds --- year 4),
for two 2-year sky maps (middle row, stars --- years 1 and 2, diamonds ---
years 3 and 4), and for the final four-year product (bottom row, diamonds).
Bottom row plots include the 68, 95, and 99\% confidence regions (heavy to
light grey) from Monte Carlo simulations of instrumental noise in the sky maps.}
\end{figure}

In this contribution I will focus on two aspects of the interpretation and
visualisation  of the \cobedmr 
data: 
1) derivation of the angular power spectrum of 
the CMB anisotropy using a data reduction method which is {\it independent} of the 
cosmological model, and
2) linear filtering of the sky maps. 
Both these goals are pursued  within the  mathematical framework for analysis of the DMR
data which was originally described in \cite{g94a}, and \cite{g94b}. 
I will briefly describe the DMR data, 
address the necessary technical points, and proceed to presentation of  the results
and discussion.

\begin{figure}[t]
\psfig{file=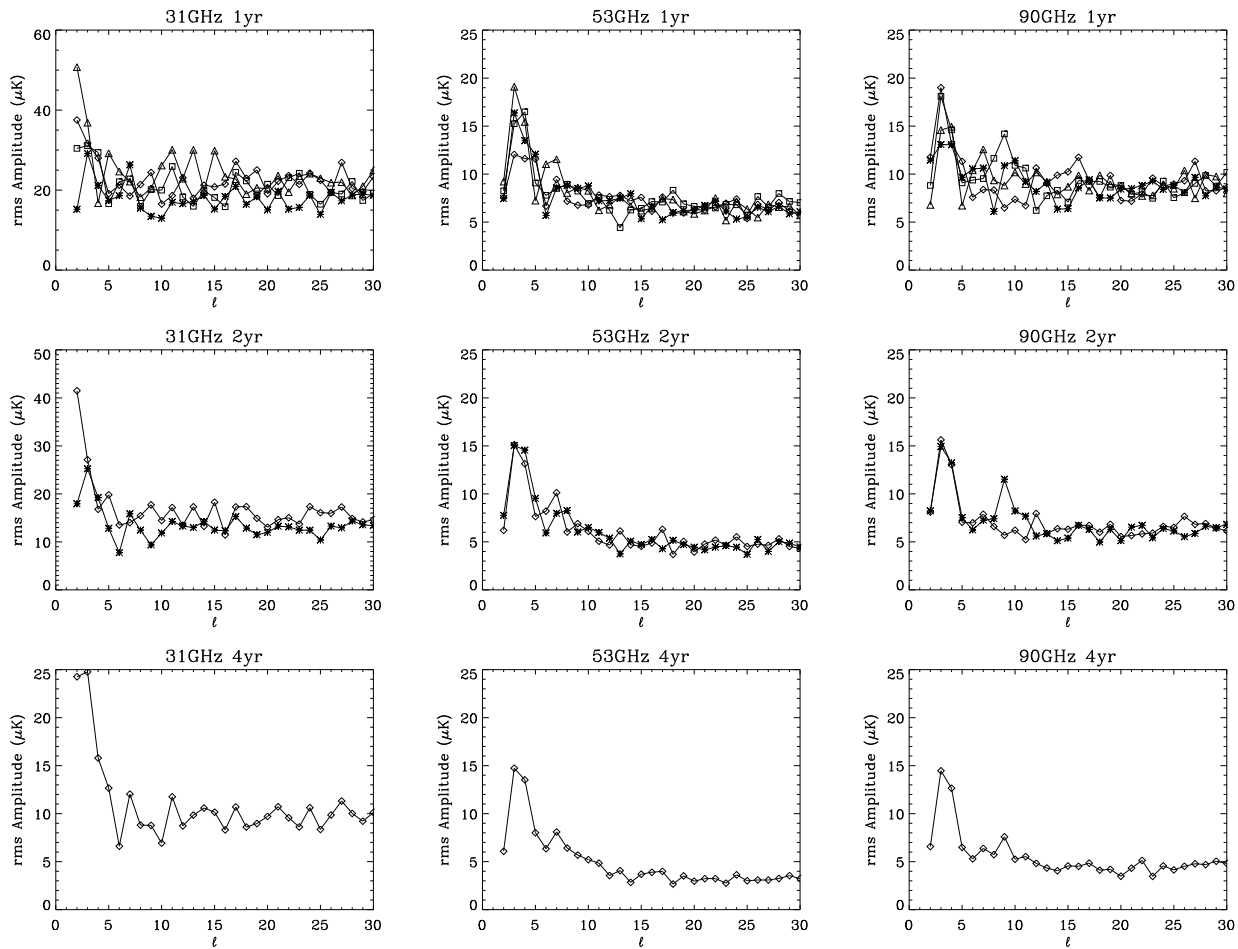,width=6.7in,angle=90}
\caption{Power spectra of the DMR (galactic frame, extended Galaxy cut,
3881 pixels) inverse-noise-variance weighted sum maps,
at each frequency of observations, 
for all  yearly, 2-year, and 4-year sky maps (same coding as in Figure 1.)
Variance weighting of the A and B sides makes the yearly and bi-yearly
effective noise levels in the 31GHz map more consistent with one another
than in the corresponding difference maps (with fifty-fifty noise contributions
from the A and B sides), and the effective noise level in the sum maps
somewhat lower than in the corresponding 
difference maps at all three frequencies.
}
\end{figure}

\section{Data}

The main result  of the \cobedmr mission  is a set of full sky maps, comprised of
two maps (A and B sides) 
at each frequency  of 31.5, 53, and 90 GHz,
pixelized in both galactic and ecliptic coordinates. 
The angular resolution of the
instrument was $\sim 7^\circ$FWHM, and the sky maps are 
divided into  6144 nearly equal area
pixels. 

A record of the sensitivity of each detector 
and the 
number of observations per pixel in each map are provided with
the sky maps. This allows for stochastic
modelling of the instrument noise, which is essential for a proper 
interpretation of the results of the data analysis
and subsequent inference of a parametric description for the cosmological signal.
[Indeed,
except for the dipole anisotropy and the galactic plane
emission, the CMB anisotropy signals 
detected  in the DMR maps are weak compared to typical noise 
contamination even at 53 GHz --- rms signal  $\sim 35\, \mu$K vs. 
rms noise $\sim 80\, \mu$K in a single pixel.]

Since the monitoring of the multichannel instrument performance was very long,
DMR is 
to date
unparalelled among the CMB anisotropy experiments in terms of noise control.

Using the customized Fourier tools introduced in \cite{g94a}
for the analysis of the incomplete sky of the Galaxy-excised
DMR sky maps one can extract from the data the cut-sky power spectra
of the sum and difference maps at each frequency, which are shown in 
Figures 1 and 2. Quantities plotted there are the rms values of the
cut-sky harmonic coefficients (the usual $(2\ell +1)$ degrees of freedom
per $\ell$-mode). 

One should appreciate from Figure 1 the steady decrease with time 
($\sim 1/\sqrt{t}$) 
of the rms noise amplitude, the excellent consistency of the yearly performances of the 
53 and 90 GHz instruments, and very good 
consistency of the measured noise properties with the 
Gaussian model.

Figure 2 shows the power spectra of signal and noise. Even the totality
of the 31.5 GHz data remains substantially noisier than the higher
frequency channels, which together with uncertainties in the  galactic
foreground contribution renders its interpretation more problematic.
Visual inspection
of the 53 and 90 GHz power spectra provides a sense of strong consistency
between these independent data sets. 

\section{Maximum likelihood estimation of the angular power spectrum of
CMB anisotropy}

A method of cut-sky Fourier decomposition and 
maximum likelihood analysis of the DMR data to infer 
the parameters of a theoretical spectrum of CMB anisotropy
was presented in \cite{g94a}. 
The application of this  formalism to derive the maximum likelihood 
estimates of the overall amplitude and shape parameters for  
power law spectrum fits to the  2-yr and 4-yr DMR data was 
given in \cite{g94b}, and \cite{g96}, respectively.
Here I demonstrate how this formalism allows one to derive a 
maximum likelihood estimate of the individual multipoles
$C_\ell=\langle \vert a_{\ell m}\vert^2\rangle = a_\ell ^2$, where
$\delta T/T(\Omega) = \sum_{\ell m} a_{\ell m} Y_{\ell m} (\Omega)$.

A proper analysis of a relatively complex data set such as  DMR
and the correct inference of a parametrisation of the sky distribution 
of
CMB anisotropy
requires the  simultaneous treatment of several interrelated problems, which include:

\noindent 1) incomplete sky coverage and resulting multipole
coupling after excision of the galactic plane region from the analysed maps;

\noindent 2) removal of the cosmologically irrelevant 
low order, $\ell = 0,\; {\rm and}\, 1$ modes,
preferably without affecting the higher order modes, which do carry  
cosmological information;

\noindent 3) modeling of the nonuniform noise distribution;

\noindent 4) identification and removal of possible diffuse foreground
emission in the areas outside the galactic cut;

\noindent 5) execution of an unbiased, or least-biased, parameter inference,
preferably using the probability density which describes the tested model
of anisotropy in an exact manner.

A program of analysis of the DMR data which addressed all of these
points was conducted in \cite{g94a},
\cite{g94b}, and \cite{g96}. An important feature of this method
of analysis is that 
it employs geometrically constructed
orthogonalized  spherical harmonics, which are {\it independent} of any
cosmological theories that one might endeavour to test with the DMR data.
This is radically different from the possible Karhunen-Lo\`eve type
methods, where one builds the basis of modes for sky map
decomposition using a specific theoretical spectrum of CMB anisotropy 
(e.g. Bond and Jaffe in this volume).

\begin{figure}[t]
\psfig{file=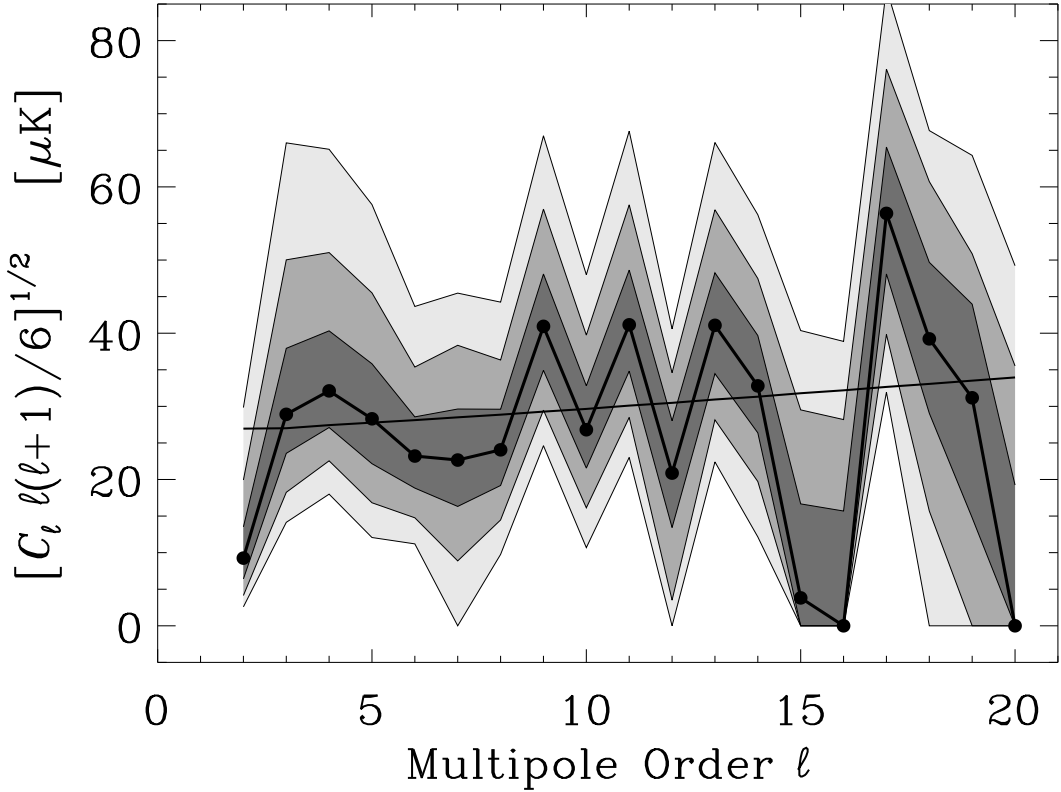,width=6.5in,angle=90}
\caption{Power spectrum of the CMB anisotropy derived by the maximum
likelihood method with each $C_\ell$ coefficient treated as a separate
parameter (hence $\Delta \ell/\ell = 0$). Noise and incomplete
sky coverage induce correlations, impossible to show on the plot, between
the derived confidence intervals on each $C_\ell$. One should note that
a correct estimator of excess variance in signal over noise in the data
was applied, hence no negative $C_\ell$ values were obtained.
A low-$\ell$ tail of the CDM model CMB anisotropy spectrum (independently
fitted to the same data) is shown as a slightly tilted solid line.}
\end{figure}

Let us now establish how the method introduced in \cite{g94a} can be 
used to compute an estimator of the full sky angular power spectrum, i.e. the  
coefficients $C_\ell$ which can be determined from the data.
Since we work with the Fourier amplitudes $\bc$ which are linear combinations
of pixel temperatures (as they are integrals on the cut-sky 
of the map products with
the cut-sky orthogonalized spherical harmonic basis functions),
their joint probability distribution is 
Gaussian (see \cite{g94a} for formal details)
\begin{equation}
P(\bc)\,d\bc = {d\bc\over (2\pi)^{N/2}\sqrt{{\rm det}(\bS+\bN)}}\, 
\exp \left(-\bc^T \cdot (\bS+\bN)^{-1} \cdot \bc \,/2\right)
\label{eq:prob},
\end{equation}
where $\bC = \bS + \bN$, and $\bS$ and $\bN$ are covariance matrices of the 
signal and noise, respectively. For the maximum likelihood calculations I use
the likelihood function defined as
\begin{equation}
F(\bA\vert \widehat\bc) = 
{\rm det}[\bS(\bA)+\bN]\, +\, \widehat\bc^T \cdot [\bS(\bA)+\bN]^{-1} \cdot 
\widehat \bc
\label{eq:likelihood},
\end{equation}
where
the full sky signal variance matrix has the form 
\begin{equation}
\bA = {\rm diag} [
\underbrace{a_2^2, \ldots, a_2^2}_{5\rm\;times},
\underbrace{a_3^2, \ldots, a_3^2}_{7\rm\;times},\, \ldots\, ,
\underbrace{a_{\ell_{max}}^2, \ldots, 
a_{\ell_{max}}^2}_{(2\ell_{max}+1)\rm\;times}]
\label{eq:varmatrix}.
\end{equation}

The maximum likelihood  problem as formulated in eqs. (1-3) ---
seeking the coefficients $C_2$-$C_{\ell_{max}}$ given the data vector 
$\widehat\bc$ --- involves the assumption of a Gaussian distribution
of both noise and CMB anisotropies, and the statistical isotropy of the
field of temperature perturbations (hence only the $\ell$ dependence
of $\langle \vert a_{\ell m}\vert^2\rangle$),
but does not imply any relation between $C_\ell$-s at different
values of $\ell$ --- hence we are solving for the
power spectrum of CMB anisotropy fully {\it indpendently} of any
specific cosmological model.

A maximum likelihood search for the power spectrum that best matches
the distribution of signal on the sky simply entails finding 
a minimum of the function $F(\bA)$ in the $\ell_{max}-1$-dimensional 
space. In the present case it is greatly facilitated by our ability
to write down easily the expression for the gradient
$\partial F/\partial a_\ell$, and implement a standard 
numerical routine for an extremum search with gradient zeroing.
The solution to this problem for the DMR 4yr galactic frame 
data is shown in Figure 3.
Our power spectrum estimator is really an estimator of excess variance
in signal over noise in the data. One should note in Fig. 3
the asymmetry of the confidence intervals around the maximum likelihood
solutions for the $C_\ell$-s, and the fact that no derived values 
go negative. If there is no signal content in the data that
can be attributed to a given  $C_\ell$, an upper limit is established,
as e.g. for $\ell =16$. The signal to noise ratio in Fourier space
drops to $\sim 1$ at $\ell \sim 15$ in the DMR 4yr data. It is 
important to see that nearly all multipoles within that
range are determined with great significance. As for the 
confidence intervals, the plot can not convey the notion of 
correlations between the errors on different multipoles. These correlations
are quite small, and more significant for multipoles separated by
$\Delta \ell =2$ than otherwise.
Because of the nontrivial (non-Gaussian) structure of confidence interval correlations
it is difficult to implement an exact method of fitting of the theoretical
model of CMB anisotropy to the DMR data by using the power spectrum,
compressed rendition of the data. For such applications the accurate linear
methods, as e.g. \cite{g94b}, should be preferred.


\section{Linear filtering of the DMR sky maps}

When the statistical properties of the signal, polluted by noise in the data,
are known, one can apply powerful tools of image processing to
improve, oftentimes visually quite dramatically, the presentation
of the measurements. 
The best known device for linear filtering of noisy data is the celebrated
Wiener filter. Alternatives, usually referred to as
sub-optimal filters, do exist, and on occasion provide a preferred
means of treating the noisy data.
In the case of the DMR raw data the signal to noise ratio is pretty poor even in
the 4yr data set. The Wiener filter, which practically ``annihilates'' those
components of the data which are assesed to be noise dominated, when
applied to  data of low quality usually
renders the impression of ``over-smoothing''. I choose to 
apply another device, called the power-equalization filter, which
is constructed as follows.

First let us define the measurement, signal, and noise vectors,
and their covariance matrices:
\begin{equation}
\bm = \bs + \bn,\;\; 
\langle \bs\cdot \bs^T \rangle =\bS,\;\; 
\langle \bn\cdot \bn^T \rangle =\bN,\;\;
\langle \bm\cdot \bm^T \rangle =\bS +\bN.
\label{eq:tofilter}
\end{equation}
Next, define the filtered data vector, and the filter matrix: 
\begin{equation}
\bof = \bF\cdot \bm = \bF\cdot(\bs + \bn).
\label{eq:filtered}
\end{equation}
If the filter is constructed from the requirement that
the mean square deviation between the filtered and the true
signal,
$\epsilon = \langle (\bof-\bs)^T\cdot(\bof-\bs)\rangle = 
{\rm tr}\langle (\bof-\bs)\cdot(\bof-\bs)^T\rangle$,
is minimized, one obtains the Wiener filter
\begin{equation}
\bF_W = \bS \cdot (\bS + \bN)^{-1}.
\label{eq:wiener}
\end{equation}

Another possibility for sub-optimal filtering, which renders the power
equalization filter, arises when
$\bF$ is chosen such that 
\begin{equation}
\langle \bof \cdot \bof^T \rangle = \bS,\;\;
{\rm i.e.}\;\;
\bF_{PE}\cdot(\bS + \bN)\cdot\bF_{PE}^T = \bS.
\label{eq:pe}
\end{equation}
Equation (7) means that on average over many applications the 
power-equalization
filter renders filtered data, whose statistical distribution matches that
of the underlying signal. 

Using Choleski decomposition of the relevant matrices, and the 
following auxiliary matrices, we can construct the power equalization filter
as follows: 
\begin{equation}
\bS^{-1}=\bL_{\bS^{-1}}\cdot\bL^T_{\bS^{-1}},\;\;\;\;\;\;
\bG_{\bS^{-1}}=\bL_{\bS^{-1}}^{-1},\;\;\;\;\;\;
\bS=\bG^T_{\bS^{-1}}\cdot\bG_{\bS^{-1}},\;\;\;\;\;\;
\label{eq:striangle}
\end{equation}
\begin{equation}
(\bS+\bN)^{-1}=\bL_{(\bS+\bN)^{-1}}\cdot\bL^T_{(\bS+\bN)^{-1}},\;\;\;
\bG_{(\bS+\bN)^{-1}}=\bL_{(\bS+\bN)^{-1}}^{-1},\;\;\;
\bS+\bN=\bG^T_{(\bS+\bN)^{-1}}\cdot\bG_{(\bS+\bN)^{-1}},\;\;
\label{eq:sstriangle}
\end{equation}
and as a result:
\begin{equation}
\bF_{PE}=\bG_{\bS^{-1}}^T\cdot\bL_{(\bS+\bN)^{-1}}^T
\label{eq:peconstr}
\end{equation}
This is an upper triangular matrix, hence filtering of a given mode
in the cut-sky data vector $\widehat \bc$
does not mix in the information from lower order modes, preserving the 
information ordering imposed by the construction of orthonormal modes
(see \cite{g94a}).
 
For the presentation purposes of this contribution I constructed separate
filters for the DMR4 53 and 90 GHz maps. The tremendous advantage of DMR
is clearly visible here. We can internally compare two
sets of decent quality observations at 53 and 90 GHz,
and we are able to quantify and 
visualize the consistency with which the instrument reveals the 
image of the last scattering surface.

\section{The first image of last scattering surface}

It should be realized that the \cobedmr 
observations 
{\it did} allow us to make a first reliable  picture
of the last scattering surface. Recognition  of this fact 
is somewhat obscured by the memory of the picture of the first DMR all sky 
map that was circulated in 1992 - 
a picture which was 
substantially distorted by noise features.

Four years after that presentation, the quality of DMR data 
is now substantially better, but it is still somewhat difficult 
to argue about its validity 
as an image of the
last scattering surface,  as
there is simply no other data at microwave frequencies
at the same angular scales
to allow a direct image to image comparison with DMR.
However, since DMR observed the sky at three frequencies,
and two of these have rendered 
sky maps of decent quality, an internal image comparison
is possible.

What I present here is a result of a fairly long chain of events:
separate instruments (with different noise properties)
were used to observe the sky (looking at different
directions at the time of observations!),
the data were separately analysed and corrected for the systematic effects,
the maps were made separately,
all the subsequent data reduction  was performed separately,
and in the end different filters were constructed and applied.

The end product of these operations is visible in two panels on the left side
of the color plate as the DMR 53 and 90 GHz, power-equalisation filtered sky 
maps. To facilitate the comparison and assessment of consistency of these maps
I add and subtract them. This is shown in the remaining panels of the 
color plate in both Mollweide and spherical polar projections.

It is immediately clear that the 53 and 90 GHz maps contain many 
common features on large angular scales (and, as shown in
\cite{extrag}, \cite{kgal}, \cite{g96}, they can not be attributed
to foreground emission from either the Galaxy, or nearby
extragalactic objects).
The epistemological value of this result of the \cobedmr mission
should not be underestimated, as this is the first time ever that
we are afforded a reliable picture of the most remote regions of the 
universe. (I shall stress that I do not attempt to make an
argument about the wonders of image processing, but about our 
truly
good luck with the DMR's splendid performance at two different frequencies!)

The rms anisotropy in the $(53+90)/2$ PE filtered map
is 35$\mu$K, while the rms in the $(53-90)/2$ map is about 15$\mu$K.
These numbers define the accuracy to which \cobedmr
has revealed to us a
most elusive astronomical realm, and  provided a first  direct glimpse at the
universe in its embryonic state.

\acknowledgements{I gratefully acknowledge the efforts of the {\it COBE} team
and the support of the Office of Space Sciences at NASA. I am grateful to
A.J. Banday for
help with preparation of the manuscript.}

%
%
\vfill

\end{document}